\documentclass[12pt]{article}
\usepackage{graphicx}
\usepackage{amsmath}
\usepackage{amsfonts}
\usepackage{amssymb}

\begin{document}

\title{\textbf{An exact macroscopic extended model with many moments}}
\author{M.C. Carrisi, S. Pennisi, A. Scanu}
\date{}
\maketitle
 \small {\em \noindent Dipartimento di Matematica ed Informatica,
Universit\`{a} di Cagliari, Via Ospedale 72,\,\ 09124 Cagliari, Italy; e-mail:
cristina.carrisi@tiscali.it;spennisi@unica.it;scanu.erice@tiscali.it}
 \\
~\\
 Extended  Thermodynamics is a very important theory: for example, it
predicts hyperbolicity, finite speeds of propagation waves as well
as continuous dependence on initial data. Therefore, it
constitutes a significative improvement of ordinary
thermo\-dy\-na\-mics. Here its methods are applied to the case of
an arbitrary, but fixed, number of moments. The kinetic approach
has already been developed in literature; then, the macroscopic
approach is here considered and the constitutive functions
appearing in the balance equations are determined up to whatever
order with respect to thermodynamical equilibrium. The results of
the kinetic approach are a particular case of the present ones.
\section{Introduction}
Extended Thermodynamics is a well established and appreciated physical theory (see refs.
\cite{1}, \cite{2} regarding the first pioneering paper on this subject and an exhaustive
description of the results which has been subsequently found). More recent results
regarding the kinetic approach are described in refs \cite{3}-\cite{6} and many
interesting properties are there obtained and exposed . The macroscopic approach has also
been investigated but the exact solution of the condition which are present in the theory
with many moments is still lacking. This gap is here filled and the general solution is
obtained.\\
The balance equations of extended thermodynamics with an arbitrary number of moments are
\begin{equation}\label{1}
\partial_{t}F_{i_{1}...i_{n}}+\partial_{k}F_{i_{1}...i_{n}k}=S_{i_{1}...i_{n}}\quad
\text{for n=0,...,N},
\end{equation}
where we call F the tensor $F_{i_{1}...i_{n}}$ when n=0.\\
Here the various tensors  are symmetric and $F_{i_{1}...i_{N}k}$ and $S_{i_{1}...i_{N}}$
are supposed to be functions of the previous one, in order to obtain a closed system. In
particular $F$, $F_i$, $F_{ll}$, $F_{ill}$ denote the densities of mass, momentum,
energy, and energy flux respectively. In this way eqs. $(\ref{1})$ for $n=0,1$, and the
trace of eqs. $(\ref{1})$ for n=2 are the conservation laws of mass, momentum and
energy; obviously to this end it is necessary to assume that S=0, $S_i=0$ and $S_{ll}=0$.\\
Eq. $(\ref{1})$ can be rewritten in a more compact form using a 4-dimensional notation in
a space that we suppose to be Euclidean (nothing will change if the space is
pseudo-Euclidean with -+++ signature, so we have chosen the simpler case).\\
In particular, let us define the symmetric tensors $M^{\alpha_{1}...\alpha_{N+1}}$ and
$S^{\alpha_{1}...\alpha_{N}}$ as follows:
\begin{enumerate}
    \item the Greek indexes go from 0 to 3,
    \item $M^{i_{1}...i_{n}0...0}=F_{i_{1}...i_{n}}$ for
    n=0,...,N+1
    \item $S^{i_{1}...i_{n}0...0}=S_{i_{1}...i_{n}}$ for
    n=0,...,N.
\end{enumerate}
In that way the balance equations $(\ref{1})$ can be simply written as
\begin{equation}\label{2}
\partial_{\alpha}M^{\alpha\alpha_{1}...\alpha_{N}}=S^{\alpha_{1}...\alpha_{N}},
\end{equation}
where $\partial_{\alpha}$ for $\alpha=0$ means the partial derivative with respect to time .\\
The entropy principle for this equations, by using Liu's theorem \cite{7} ensures the
existence of the parameters $L_{\alpha_1\cdots\alpha_N}$, called Lagrange Multipliers,
such that
\begin{equation}\label{2'}
dH^{\alpha}=L_{\alpha_1\cdots\alpha_N}dM^{\alpha_1\cdots\alpha_N\alpha}, \quad
L_{\alpha_1\cdots\alpha_N}S^{\alpha_1\cdots\alpha_N}\geq 0
\end{equation}
where $H^0$ is the entropy density and $H^i$ its flux. A brilliant Ruggeri's idea is to
define
\begin{equation}\label{2''}
H^{'\alpha}=-H^{\alpha}+L_{\alpha_1\cdots\alpha_N}M^{\alpha_1\cdots\alpha_N\alpha}
\end{equation}
and to take the Lagrange Multipliers as independent variables. In this way eq.
$(\ref{2'})_1$ becomes $d
H^{'\alpha}=M^{\alpha_1\cdots\alpha_N\alpha}dL_{\alpha_1\cdots\alpha_N}$, from which
\begin{equation}\label{3}
M^{\alpha_{1}\alpha_{2}...\alpha_{N+1}}=\frac{\partial H^{'\alpha_{N+1}}}{\partial
L_{\alpha_{1}...\alpha_{N}}}
\end{equation}
In this way the tensors appearing in the balance equations $(\ref{2})$ are found as
functions of the parameters $L_{\alpha_{1}...\alpha_{N}}$, called also mean field, as
soon as $H^{'\alpha}$ is known. Obviously $L_{\alpha_{1}...\alpha_{N}}$ is symmetric. By
substituting $(\ref{3})$ into eq. $(\ref{2})$ this takes the symmetric form
\begin{equation*}
\frac{\partial^2 H^{'\alpha_{N+1}}}{\partial L_{\beta_1\cdots\beta_N}\partial
L_{\alpha_1\cdots\alpha_N}}\partial_{\alpha_{N+1}}L_{\beta_1\cdots\beta_N}=S^{\alpha_1\cdots\alpha_N},
\end{equation*}
so that hyperbolicity is ensured provided that $H^{'\alpha}$ is a convex function of the
mean field. By eliminating these parameters from eqs. $(\ref{3})$ we obtain
$F_{i_{1}...i_{N+1}}$ again, as function of F, $F_{i}$, ..., $F_{i_{1}...i_{N}}$. If we
want a model in which some among eqs. $(\ref{1})$ is present only by means of one of
its traces, it can be obtained from the present model with the method of the subsystems \cite{2}.\\
Note that eq. $(\ref{3})$ for
$\alpha_{1}\alpha_{2}...\alpha_{N+1}=i_{1}...i_{n}i_{n+1}0...0$ and for
$\alpha_{1}\alpha_{2}...\alpha_{N+1}=i_{1}...i_{n}0...0i_{n+1}$ gives respectively
\begin{equation}\label{4}
F_{i_{1}...i_{n}i_{n+1}}=\frac{\partial H^{'0}}{L_{i_{1}...i_{n}i_{n+1}}}\qquad , \qquad
F_{i_{1}...i_{n}i_{n+1}}=\frac{\partial H^{'n+1}}{L_{i_{1}...i_{n}}}
\end{equation}
as in the 3-dimensional notation.\\
So, to impose eq. $(\ref{3})$ we have to find the more general expression of
$H^{'\alpha}$ such that $M^{\alpha_{1}\alpha_{2}...\alpha_{N+1}}$ is symmetric. We will
refer to this as ``the symmetry condition". We will impose also the principle of galilean
invariance; this has been exploited in \cite{2},\cite{8}-\cite{10} for a generic system
of balance laws. In section 2 we will apply these results to our system, taking care of
converting
them in the present 4-dimensional notation, so obtaining further conditions. \\
In section 3 these, together with the symmetry condition, will be investigated and their
solution will be found up to whatever order with respect to thermodynamical equilibrium,
except for two numerable families of constants arising from integration. In section 4 it
will be shown that the results of the kinetic approach are a particular case of the
present one so that, as usual, the macroscopic approach is more general than the kinetic
one.  If we rewrite the present paper with N-1 instead of N, we find the model with less
moments with the method used in this work, which we call the ``direct method". But in
ref. \cite{2} it has been shown that a solution (for the model with N-1 instead of N) can
be obtained also with the ``method of subsystems"; this consists in taking the
costitutive functions of the model with N as maximum order of moments, and in calculating
them in $\lambda_{i_1\cdots i_N=0}$, i.e., for zero value of the 3-dimensional Lagrange
multiplier with the greatest order. In section 5 we will see that the solution obtained
with the method of subsystems is a particular one of that  obtained with the direct
method; more explicitly, it can be obtained from the latter by considering equal to zero
one of the above mentioned families of constants. At last, conclusions will be drown.
\section{The Galilean relativity principle.}
To impose this principle, it is firstly necessary to know how our variables transform
under a change of galileanly equivalent frames $\Sigma$ and $\Sigma'$. This problem has
been studied by Ruggeri in \cite{8} and we have only to write its results in our
4-dimensional form. This is easily achieved in the kinetic model because the kinetic
counterpart of $M^{\alpha_1\cdots\alpha_{N+1}}$ is
\begin{equation}\label{5}
M^{\alpha_1\cdots\alpha_{N+1}}=\int f c^{\alpha_1}\cdots c^{\alpha_{N+1}} d\underline{c}
\end{equation}
with $c^0=1$, $d\underline{c}=dc^1 dc^2 dc^3$ and f is the distribution function.
Consequently, in $\Sigma'$ we have
\begin{equation*}
m^{\alpha_1\cdots\alpha_{N+1}}=\int f c^{'\alpha_1}\cdots c^{'\alpha_{N+1}} dc'
\end{equation*}
and, if $v^i$ is the constant velocity of each point of $\Sigma'$ with respect to
$\Sigma$, we have $c^{\alpha}=c^{'\alpha}+v^{\alpha}$, with $v^0=0$. It follows that
\begin{equation*}
M^{\alpha_{1}...\alpha_{N+1}}=\sum_{i=0}^{N+1} \left(
\begin{array}{c}
  N+1 \\
  i \\
\end{array}
\right) v^{(\alpha_{1}}...v^{\alpha_{i}}m^{\alpha_{i+1}...\alpha_{N+1})}
\end{equation*}
or
\begin{equation}\label{6}
M^{\alpha_{1}...\alpha_{N+1}}=\sum_{i=0}^{N+1} \left(
\begin{array}{c}
  N+1 \\
  i \\
\end{array}
\right)
v^{(\alpha_{1}}...v^{\alpha_{i}}m^{\alpha_{i+1}...\alpha_{N+1})\beta_{1}...\beta_{i}}t_{\beta_{1}}...t_{\beta_{i}}
\end{equation}
with $t_{\mu}\equiv(1,0,0,0)$ for our previous notation. We obtain the transformation of
$M^{\alpha_{1}...\alpha_{N}0}$ (which was the initial independent variable) multiplying
eq. $(\ref{6})$ by $t_{\alpha_{N+1}}$ so finding
\begin{equation}\label{7}
M^{\alpha_{1}...\alpha_{N}0}=X^{\alpha_{1}...\alpha_{N}}_{\beta_{1}...\beta_{N}}(\underline{v})m^{\beta_{1}...\beta_{N}0}
\end{equation}
with
\begin{equation}\label{8}
X^{\alpha_{1}...\alpha_{N}}_{\beta_{1}...\beta_{N}}=\sum_{i=0}^{N}\left(%
\begin{array}{c}
  N \\
  i \\
\end{array}%
\right)t_{(\beta_{1}}...t_{\beta_{i}}v^{(\alpha_{1}}...v^{\alpha_{i}}\delta^{\alpha_{i+1}}_{\beta_{i+1}}...\delta^{\alpha_{N)}}_{\beta_{N)}}
\end{equation}
where we have taken into account of $v^{0}=0$, of the identity $\begin{pmatrix}
  N+1 \\
i
\end{pmatrix}\frac{N+1-i}{N+1}=\begin{pmatrix}
 N \\
  i
\end{pmatrix}$ and that the term with i=N+1 gives a null contribution. Comparison between $(\ref{7})$
and $(\ref{8})$ with $(\ref{6})$ shows that
$X^{\alpha_{1}...\alpha_{N}}_{\beta_{1}...\beta_{N}}$ could be obtained from
$X^{\alpha_{1}...\alpha_{N+1}}_{\beta_{1}...\beta_{N+1}}$ simply replacing N+1 with N.
From eq. $(\ref{8})$ it follows also
\begin{equation}\label{8''}
X^{\alpha_{1}...\alpha_{N}\alpha}_{\beta_{1}...\beta_{N}\beta}=
X^{\alpha_{1}...\alpha_{N}}_{(\beta_{1}...\beta_{N}}t_{\beta)}v^{\alpha}+
X^{\alpha_{1}...\alpha_{N}}_{(\beta_{1}...\beta_{N}}\delta^{\alpha}_{\beta)}.
\end{equation}
Similarly, $H^{\alpha}$ transforms according to the rule
\begin{equation}\label{8'}
H^{\alpha}=h^0 v^{\alpha}+h^{\alpha},
\end{equation}
of which $H^0=h^0$ is a component.\\
Eqs. $(\ref{6})$ and $(\ref{8'})$ have been obtained with the kinetic model only for the
sake of simplicity; it is obvious that they hold also in the macroscopic case. The
transformation rule of the Lagrange multipliers can be obtained now from $(\ref{2'})_1$
with $\alpha=0$, i.e.
\begin{equation*}
dh^0=dH^0=L_{\alpha_1\cdots\alpha_N}dM^{\alpha_1\cdots\alpha_N 0}=
L_{\alpha_1\cdots\alpha_N} X^{\alpha_{1}...\alpha_{N}}_{\beta_{1}...\beta_{N}}
dm^{\beta_1\cdots\beta_N 0}
\end{equation*}
where $(\ref{8'})$ and $(\ref{7})$ have been used. In other words we have
\begin{equation}\label{8*}
dh^0=l_{\beta_1\cdots\beta_N}dm^{\beta_1\cdots\beta_N 0}
\end{equation}
with
\begin{equation*}
l_{\alpha_{1}...\alpha_{N}}=X_{\alpha_{1}...\alpha_{N}}^{\beta_{1}...\beta_{N}}L_{\beta_{1}...\beta_{N}}
\end{equation*}
i.e.
\begin{equation}\label{9}
l_{\alpha_{1}...\alpha_{N}}=\sum_{i=0}^{N}\left(%
\begin{array}{c}
  N \\
  i \\
\end{array}%
\right)t_{(\alpha_{1}}...t_{\alpha_{i}}v^{\beta_{1}}...v^{\beta_{i}}L_{\alpha_{i+1}...\alpha_{N)}\beta_{1}...\beta_{i}}.
\end{equation}
A consequence of this result can be obtained from $(\ref{2''})$ with $\alpha=0$ and
written in the frame $\Sigma'$, i.e.,
$h^{'0}=-h^{0}+l_{\alpha_1\cdots\alpha_N}m^{\alpha_1\cdots\alpha_N 0}$; it follows $d
h^{'0}=m^{\alpha_1\cdots\alpha_N 0}d l_{\alpha_1\cdots\alpha_N}$ from which
\begin{equation}\label{8**}
m^{\alpha_1\cdots\alpha_N 0}=\frac{\partial h^{'0}}{\partial l_{\alpha_1\cdots\alpha_N}}
\end{equation}
as in $\Sigma$. Moreover, from $(\ref{2''})$, $(\ref{8'})$, $(\ref{6})$, $(\ref{8''})$,
$(\ref{9})$ and again $(\ref{2''})$ and $(\ref{8'})$ it follows
\begin{eqnarray*}
H^{'\alpha}&=&-h^0v^{\alpha}-h^{\alpha}+L_{\alpha_1\cdots\alpha_N}
X^{\alpha_{1}...\alpha_{N}\alpha}_{\beta_{1}...\beta_{N}\beta}m^{\beta_1\cdots\beta_N\beta}\\
&=& -h^0v^{\alpha}-h^{\alpha}+l_{\beta_1\cdots\beta_N}m^{\beta_1\cdots\beta_N
0}v^{\alpha}+ l_{\beta_1\cdots\beta_N}m^{\beta_1\cdots\beta_N\alpha}
\end{eqnarray*}
i.e.,
\begin{equation}\label{10}
H^{'\alpha}=h^{'0}v^{\alpha}+h^{'\alpha}
\end{equation}
which is similar to $(\ref{8'})$.\\
We are now ready to consider the Galilean relativity principle. It imposes that the
following diagram is commutative\\
\begin{picture}(200,190)(0,120)
\put(2,280){\framebox(50,25){$L_{\alpha_1\cdots\alpha_N}$}}
\put(250,280){\framebox(150,25){$l_{\gamma_1\cdots\gamma_N}=X^{\alpha_{1}...\alpha_{N}}_{\gamma_{1}...\gamma_{N}}L_{\alpha_1\cdots\alpha_N}$}}
\put(2,200){\framebox(100,50)}
\put(5,230){\mbox{$M^{\beta_1\cdots\beta_{N+1}}(L_{\alpha_1\cdots\alpha_N})$}}
\put(5,210){\mbox {$H^{'\alpha}(L_{\alpha_1\cdots\alpha_N})$}}
\put(2,120){\framebox(190,50)}
\put(5,150){\mbox{$X^{\beta_{1}...\beta_{N+1}}_{\delta_{1}...\delta_{N+1}}m^{\delta_1\cdots\delta_{N+1}}(X^{\alpha_{1}...\alpha_{N}}_{\gamma_{1}...\gamma_{N}}L_{\alpha_1\cdots\alpha_N})$}}
\put(5,130){\mbox{$v^{\alpha}t_{\delta}h^{'\delta}(\cdots)+h^{'\alpha}(\cdots)$}}
\put(250,120){\framebox(140,50)}
\put(255,150){\mbox{$m^{\delta_1\cdots\delta_{N+1}}(X^{\alpha_1\cdots\alpha_N}_{\gamma_1\cdots\gamma_N}L_{\alpha_1\cdots\alpha_N})$}}
\put(255,130){\mbox{$h^{'\delta}(X^{\alpha_1\cdots\alpha_N}_{\gamma_1\cdots\gamma_N}L_{\alpha_1\cdots\alpha_N})$}}
\put(68,290){\vector(1,0){160}} \put(25,275){\vector(0,-1){20}}
\put(330,270){\vector(0,-1){90}} \put(25,190){\line(0,-1){10}}
\put(30,190){\line(0,-1){10}} \put(245,147){\vector(-1,0){50}}
\end{picture}\\
In other words, we must have
\begin{eqnarray}\label{10'}
H^{'\alpha}(L_{\alpha_1\cdots\alpha_N})&=&v^{\alpha}t_{\delta}h^{'\delta}
(X^{\alpha_{1}...\alpha_{N}}_{\gamma_{1}...\gamma_{N}}(\underline{v})
L_{\alpha_1\cdots\alpha_N})+
h^{'\alpha}(X^{\alpha_{1}...\alpha_{N}}_{\gamma_{1}...\gamma_{N}}(\underline{v})
L_{\alpha_1\cdots\alpha_N}) \nonumber\\
M^{\beta_1\cdots\beta_{N+1}}(L_{\alpha_1\cdots\alpha_N})&=&
X^{\beta_{1}...\beta_{N+1}}_{\delta_{1}...\delta_{N+1}}(\underline{v})
m^{\delta_1\cdots\delta_{N+1}}(
X^{\alpha_{1}...\alpha_{N}}_{\gamma_{1}...\gamma_{N}}(\underline{v})
L_{\alpha_1\cdots\alpha_N})
\end{eqnarray}
Eq. $(\ref{10'})_2$, by using eqs. $(\ref{8''})$ and $(\ref{8**})$ becomes
\begin{eqnarray*}
M^{\beta_1\cdots\beta_N\alpha}&=& X^{\beta_{1}...\beta_{N}}_{\delta_{1}...\delta_{N}}
m^{\delta_1\cdots\delta_N 0}v^{\alpha}+
X^{\beta_{1}...\beta_{N}}_{\delta_{1}...\delta_{N}}m^{\delta_1\cdots\delta_N\alpha}=\\
&=&X^{\beta_{1}...\beta_{N}}_{\delta_{1}...\delta_{N}}\frac{\partial h^{'0}}{\partial
l_{\delta_1\cdots\delta_N}}v^{\alpha}+X^{\beta_{1}...\beta_{N}}_{\delta_{1}...\delta_{N}}m^{\delta_1\cdots\delta_n\alpha}
\end{eqnarray*}
Now the derivative of $(\ref{10'})_1$ with respect to $L^{\beta_1\cdots\beta_N}$ is
\begin{equation*}
\frac{\partial H^{'\alpha}}{\partial
L^{\beta_1\cdots\beta_N}}=M^{\beta_1\cdots\beta_N\alpha}=v^{\alpha}\frac{\partial
h^{'0}}{\partial
l_{\gamma_1\cdots\gamma_N}}X^{\beta_1\cdots\beta_N}_{\gamma_1\cdots\gamma_N}(\underline{v})+
\frac{\partial h^{'\alpha}}{\partial
l_{\gamma_1\cdots\gamma_N}}X^{\beta_1\cdots\beta_N}_{\gamma_1\cdots\gamma_N}(\underline{v}).
\end{equation*}
It follows that eq. $(\ref{10'})_2$, holds iff
\begin{equation*}
\frac{\partial h^{'\alpha}}{\partial
l_{\gamma_1\cdots\gamma_N}}X^{\beta_1\cdots\beta_N}_{\gamma_1\cdots\gamma_N}=
m^{\delta_1\cdots\delta_N\alpha}X^{\beta_1\cdots\beta_N}_{\gamma_1\cdots\gamma_N}
\end{equation*}
i.e.
\begin{equation}\label{10''}
m^{\gamma_1\cdots\gamma_N\alpha}=\frac{\partial h^{'\alpha}}{\partial
l_{\gamma_1\cdots\gamma_N}}
\end{equation}
which is the counterpart of eq. $(\ref{3})$ in the frame $\Sigma'$.\\
There remain to impose eq. $(\ref{10'})_1$. Now it becomes an identity when calculated in
$\underline{v}=0$ (see eqs. $(\ref{10})$ and $(\ref{8})$ to this regard) so that it holds
iff its derivative with respect to $v_j$ is satisfied, i.e.,
\begin{eqnarray}\label{10'''}
&&0=\frac{\partial h^{'0}}{\partial l_{\gamma_1\cdots \gamma_N}}\frac{\partial
l_{\gamma_1\cdots\gamma_N}}{\partial v_j}\qquad \qquad \quad \text{for }\alpha=0, \nonumber \\
&&0=h^{'0}\delta^{\alpha}_{j}+\frac{\partial h^{'\alpha}}{\partial l_{\gamma_1\cdots
\gamma_N}}\frac{\partial l_{\gamma_1\cdots\gamma_N}}{\partial v_j}\quad \text{ for
}\alpha=0,1,2,3.
\end{eqnarray}
The second of this has been obtained by taking into account also eq. $(\ref{10'''})_1$;
on the other hand, this is included in $(\ref{10'''})_2$ with $\alpha=0$. Eq.
$(\ref{10'''})_2$, by using eq. $(\ref{9})_{2}$ now becomes
\begin{equation*}
h^{'0}\delta^{\alpha}_j+ \frac{\partial h^{'\alpha}}{\partial
l_{\alpha_{1}...\alpha_{N}}}\sum_{i=1}^{N}\left(
\begin{array}{c}
  N \\
  i \\
\end{array}
\right) i \cdot
t_{(\alpha_{1}}...t_{\alpha_{i}}v^{\beta_{1}}...v^{\beta_{i-1}}L_{\alpha_{i+1}...\alpha_{N})\beta_{1}...\beta_{i-1}j}=0.
\end{equation*}
We remove the symmetrization with respect to $\alpha_1\cdots\alpha_N$ which is not
necessary because of the contraction with $\frac{\partial h^{'\alpha}}{\partial
l_{\alpha_{1}...\alpha_{N}}}$ which is symmetric; for the same reason we can exchange
$\alpha_{i}$ and $\alpha_{N}$ and then reintroduce the symmetrization with respect to
$\alpha_{1}...\alpha_{N-1}$, obtaining so
\begin{equation*}
h^{'0}\delta^{\alpha}_j+ t_{\alpha_{N}}\frac{\partial h^{'\alpha}}{\partial
l_{\alpha_{1}...\alpha_{N}}}\sum_{i=1}^{N}\left(
\begin{array}{c}
  N \\
  i \\
\end{array}
\right)i \cdot
t_{(\alpha_{1}}...t_{\alpha_{i-1}}v^{\beta_{1}}...v^{\beta_{i-1}}L_{\alpha_{i}...\alpha_{N-1})\beta_{1}...\beta_{i-1}j}=0.
\end{equation*}
We replace i with i+1 and we have
\begin{eqnarray}\label{11}
&& h^{'0}\delta^{\alpha}_j+\frac{\partial h^{'\alpha}}{\partial
l_{\alpha_{1}...\alpha_{N}}}t_{\alpha_{N}}\sum_{i=0}^{N-1}\left(
\begin{array}{c}
  N \\
  i+1 \\
\end{array}
\right)(i+1)\cdot
t_{(\alpha_{1}}...t_{\alpha_{i}}v^{\beta_{1}}...v^{\beta_{i}}L_{\alpha_{i+1}...\alpha_{N-1})\beta_{1}...\beta_{i}j}=0 \quad \text{, or}\nonumber\\
&& h^{'0}\delta^{\alpha}_j+\frac{\partial h^{'\alpha}}{\partial
l_{\alpha_{1}...\alpha_{N}}}t_{\alpha_{N}}\sum_{i=0}^{N-1}N\left(
\begin{array}{c}
  N-1 \\
  i \\
\end{array}
\right)\cdot
t_{(\alpha_{1}}...t_{\alpha_{i}}v^{\beta_{1}}...v^{\beta_{i}}L_{\alpha_{i+1}...\alpha_{N-1})\beta_{1}...\beta_{i}j}=0.
\end{eqnarray}
But, by using eq. $(\ref{9})$ we have
\begin{eqnarray}\label{12}
l_{\alpha\cdots\alpha_{N-1}j}&=&\sum_{i=1}^{N}\frac{i}{N}\begin{pmatrix}
  N\\
  i
\end{pmatrix} t_j t_{(\alpha_1}\cdots t_{\alpha_{i-1}}v^{\beta_1}\cdots v^{\beta_i}L_{\alpha_i\cdots \alpha_{N-1})\beta_1\cdots
\beta_i}+ \nonumber \\
&+&\sum_{i=0}^{N-1}\frac{N-i}{N}\begin{pmatrix}
  N\\
  i
\end{pmatrix} t_{(\alpha_1}\cdots t_{\alpha_{i}}v^{\beta_1}\cdots v^{\beta_i}L_{\alpha_{i+1}\cdots \alpha_{N-1})j \beta_1\cdots
\beta_i}\nonumber \\
&=& \sum_{i=0}^{N-1}\begin{pmatrix}
  N-1\\
  i
\end{pmatrix} t_{(\alpha_1}\cdots t_{\alpha_{i}}v^{\beta_1}\cdots v^{\beta_i}L_{\alpha_{i+1}\cdots \alpha_{N-1})j \beta_1\cdots
\beta_i}
\end{eqnarray}
because $t_j=0$. This allows to rewrite eq. $(\ref{11})$ as
\begin{equation}\label{13}
0=h^{'\mu}t_{\mu}\delta_{j}^{\alpha}+N\frac{\partial h^{'\alpha}}{\partial
l_{\alpha_1...\alpha_N}}t_{\alpha_N}l_{\alpha_1...\alpha_{N-1}j}.
\end{equation}
Until now we have  obtained that the entropy principle jointly with the galilean
relativity principle amounts to say that
\begin{enumerate}
  \item eqs. $(\ref{3})$ are invariant under changes of galileanly equivalent observers
  (see eq. $(\ref{10''})$),
  \item the further condition $(\ref{13})$ must hold.
\end{enumerate}
For the sake of completeness, we note that eq. $(\ref{10'})_1$ might be satisfied also
with $H^{\alpha}$ and $h^{\alpha}$, i.e.
\begin{equation*}
H^{\alpha}(L_{\alpha_1\cdots\alpha_N})=v^{\alpha}t_{\delta}h^{\delta}(
X^{\alpha_1\cdots\alpha_N}_{\gamma_1\cdots\gamma_N}L_{\alpha_1\cdots\alpha_N})+
h^{'\alpha}(
X^{\alpha_1\cdots\alpha_N}_{\gamma_1\cdots\gamma_N}L_{\alpha_1\cdots\alpha_N}).
\end{equation*}
But this is a consequence of $(\ref{10'})$ as it can be seen running over backwards the
above passages which allowed to obtain eq. $(\ref{10})$ from eq. $(\ref{8'})$. Moreover,
in \cite{9} and \cite{10} it has be proved that the conditions here obtained are the same
of the following approach:
\begin{enumerate}
  \item consider eqs. $(\ref{6})$, $(\ref{8'})$ and $(\ref{10})$ but with
  $v_i=\frac{F_i}{F}$, instead of an arbitrary constant $v_i$; in this way
  $m^{\alpha_1\cdots\alpha_{N+1}}$, $h^{\alpha}$ and $h^{'\alpha}$ become the non-convective
  parts of $M^{\alpha_1\cdots\alpha_{N+1}}$, $H^{\alpha}$ and $H^{'\alpha}$,
  respectively,
  \item impose the conditions $(\ref{10''})$ and $(\ref{13})$ but considering
  $l_{\gamma_1\cdots\gamma_N}$ independent variables,
  \item consider eqs. $(\ref{10''})$ with $\alpha=0$ and $m^{i 0\cdots 0}$ as definition of
  $l_{\gamma_1\cdots\gamma_N}=l_{\gamma_1\cdots\gamma_N}(m^{\alpha_1\cdots\alpha_N 0})$,
  and substitute this in the expressions of $m^{i_1\cdots i_{N+1}}$, $h^{\alpha}$ and
  $h^{'\alpha}$ so obtaining the closure in terms of the non-convective quantities $m^{\alpha_1\cdots\alpha_N
  0}$.
\end{enumerate}
In any case, we have to impose $(\ref{10''})$ and $(\ref{13})$; in other words we have to
find the quadrivector $h^{'\alpha_{N+1}}$ such that the right hand side of eq.
$(\ref{10''})$ is symmetric and for which eq. $(\ref{13})$ holds; after that eq.
$(\ref{10''})$ gives $m^{\beta_1\cdots\beta_N\beta_{N+1}}$. In this way we will find the
required closure satisfying the entropy principle and that of galilean relativity. This
will be done in the next section.
\section{Exploitation of the conditions $(\ref{10''})$ and $(\ref{13})$.}
We want now to impose eqs. $(\ref{10''})$ and $(\ref{13})$ up to whatever order with
respect to thermodynamical equilibrium. This is defined as the state where
\begin{equation}\label{19z}
l_{\beta_1\cdots\beta_N}=\lambda t_{\beta_1}\cdots t_{\beta_N} +\frac{1}{3}\lambda_{ll}
h_{(\beta_1\beta_2}t_{\beta_3}\cdots t_{\beta_N)}
\end{equation}
holds, with
$h_{\beta\gamma}=\delta_{\beta\gamma}-t_{\beta}t_{\gamma}=diag(0,1,1,1)$,
\begin{equation}\label{22a}
\lambda= t^{\beta_1} \cdots t^{\beta_N} l_{\beta_1\cdots\beta_N} \quad \quad
 \lambda_{ll}= \begin{pmatrix}
  N \\
  2
\end{pmatrix}h^{\beta_1\beta_2}t^{\beta_3} \cdots t^{\beta_N} l_{\beta_1 \cdots \beta_N}.
\end{equation}
We can consider the Taylor expansion for $h^{'\alpha}$
\begin{equation}\label{19}
h^{'\alpha}=\sum_{k=0}^{\infty}\frac{1}{k!}A^{\alpha B_1\cdots B_k}\tilde{l}_{B_1} \cdots
\tilde{l}_{B_k},
\end{equation}
with
\begin{equation}\label{22b}
\tilde{l}_{\beta_1\cdots \beta_N}=l_{\beta_1\cdots \beta_N}-\lambda t_{\beta_1} \cdots
t_{\beta_N}-\frac{1}{3}\lambda_{ll}h_{(\beta_1\beta_2}t_{\beta_3}\cdots t_{\beta_N)},
\end{equation}
\begin{equation}\label{19a}
A^{\alpha B_1\cdots B_k}= \left(\frac{\partial^k h{'^\alpha}}{\partial l_{B_1} \cdots
\partial l_{B_k} }\right)_{eq}
\end{equation}
where the multi-index notation $B_i=\beta_i^1 \cdots \beta_i^N $ has been used. Thanks to
eq. $(\ref{10''})$ we can exchange $\alpha$ with each other index taken from those
included in any $B_i$. So it is possible to exchange every index with all the others,
i.e., $A^{\alpha B_1\cdots B_k}$ is symmetric with respect to any couple of indexes. We
note that there are 2 compatibility conditions between eqs. $(\ref{19})$ and
$(\ref{19a})$; they can be obtained as follows: let us consider the tensor
$\frac{\partial^k h^{'\alpha}}{\partial l_{B_1} \cdots l_{B_k}}$ as function of
$\tilde{l}_B, \lambda, \lambda_{ll}$, and take the derivatives with respect to
$l_{\beta_1\cdots \beta_N}$, calculating the result at equilibrium; we find
\begin{eqnarray*}
&& A^{\alpha B_1\cdots B_k \beta_1\cdots \beta_N}= \left(\frac{\partial^{k+1}
h^{'\alpha}}{\partial l_{B_1} \cdots \partial l_{B_k}\partial\tilde{l}_{\gamma_1 \cdots
\gamma_N}} \right)_{eq} \frac{\partial\tilde{l}_{\gamma_1 \cdots \gamma_N}}{\partial
l_{\beta_1 \cdots
\beta_N}}+ \\
&& +\left(\frac{\partial^{k+1} h^{'\alpha}}{\partial l_{B_1} \cdots\partial
l_{B_k}\partial \lambda}\right)_{eq}\frac{\partial\lambda}{\partial l_{\beta_1\cdots
\beta_N}}+ \left(\frac{\partial^{k+1} h^{'\alpha}}{\partial l_{B_1} \cdots\partial
l_{B_k}\partial \lambda_{ll} }\right)_{eq}\frac{\partial\lambda_{ll}}{\partial
l_{\beta_1\cdots \beta_N}}.
\end{eqnarray*}
If we multiply this by $t_{\beta_1} \cdots t_{\beta_N}$ and by
$h_{\beta_1\beta_2}t_{\beta_3}\cdots t_{\beta_N}$ we find, respectively
\begin{equation}\label{20}
  \begin{cases}
   A^{\alpha B_1\cdots B_k\beta_1\cdots \beta_N}t_{\beta_1} \cdots t_{\beta_N} =
    \frac{\partial}{\partial \lambda}A^{\alpha B_1\cdots B_k}\\
A^{\alpha B_1\cdots B_k\beta_1\cdots \beta_N}h_{\beta_1\beta_2}t_{\beta_3}\cdots
t_{\beta_N}=3\frac{\partial}{\partial \lambda_{ll}}A^{\alpha B_1\cdots B_k},
  \end{cases}
\end{equation}
where we have taken into account that from eqs. $(\ref{22a})$ and $(\ref{22b})$ it
follows
\begin{eqnarray*}
&& \frac{\partial\lambda}{\partial l_{\beta_1 \cdots \beta_N}}= t^{\beta_1} \cdots
t^{\beta_N} ~~~~~~~~~~~~~~~~ \frac{\partial\lambda_{ll}}{\partial l_{\beta_1 \cdots
\beta_N} }= \begin{pmatrix}
  N \\
  2
\end{pmatrix}h^{(\beta_1\beta_2}t^{\beta_3} \cdots t^{\beta_N)}\\
&& \frac{\partial\tilde{l}_{\gamma_1\cdots \gamma_N}}{\partial l_{\beta_1 \cdots
\beta_N}}=g^{(\beta_1}_{\gamma_1}\cdots g^{\beta_N)}_{\gamma_N}- t^{\beta_1} \cdots
t^{\beta_N} t_{\gamma_1} \cdots t_{\gamma_N}-\frac{1}{3}\begin{pmatrix}
  N \\
  2
\end{pmatrix}h^{(\beta_1\beta_2}t^{\beta_3}\cdots t^{\beta_N)}h_{(\gamma_1\gamma_2}t_{\gamma_3}\cdots
t_{\gamma_N)},
\end{eqnarray*}
from which
\begin{eqnarray*}
&&\frac{\partial\lambda}{\partial l_{\beta_1 \cdots \beta_N}} t_{\beta_1} \cdots
t_{\beta_N}=1
\quad \quad \frac{\partial\lambda}{\partial l_{\beta_1 \cdots \beta_N}} h_{\beta_1\beta_2}t_{\beta_3} \cdots t_{\beta_N}=0\\
&&\frac{\partial\lambda_{ll}}{\partial l_{\beta_1\cdots\beta_N} }t_{\beta_1} \cdots
t_{\beta_N}=0\quad \quad \frac{\partial\lambda_{ll}}{\partial l_{\beta_1 \cdots \beta_N}}
h_{\beta_1\beta_2}t_{\beta_3} \cdots t_{\beta_N}=3
\\ &&\frac{\partial\tilde{l}_{\gamma_1\cdots \gamma_N}}{\partial l_{\beta_1\cdots\beta_N}} t_{\beta_1} \cdots t_{\beta_N} =0\quad \quad
\frac{\partial\tilde{l}_{\gamma_1\cdots \gamma_N}}{\partial l_{\beta_1 \cdots
\beta_N}}h_{\beta_1\beta_2}t_{\beta_3}\cdots t_{\beta_N}=0.
\end{eqnarray*}
It will be useful in the sequel to note a consequence of the condition $(\ref{20})$. By
using also eq. $(\ref{19})$ we have
\begin{eqnarray*}
&&  \frac{\partial h^{'\alpha}}{\partial l_{\beta_1 \cdots
\beta_N}}=\sum_{k=1}^{\infty}\frac{1}{(k-1)!}
A^{\alpha B_1\cdots B_{k-1}\gamma_1\cdots\gamma_N}\tilde{l}_{B_1}\cdots \tilde{l}_{B_{k-1}} \\
&&\left(g^{(\beta_1}_{\gamma_1}\cdots g^{\beta_N)}_{\gamma_N}- t^{\beta_1} \cdots
t^{\beta_N} t_{\gamma_1} \cdots t_{\gamma_N}-\frac{1}{3}\begin{pmatrix}
  N \\
  2
\end{pmatrix}h^{(\beta_1\beta_2}t^{\beta_3}\cdots t^{\beta_N)}h_{(\gamma_1\gamma_2}t_{\gamma_3}\cdots t_{\gamma_N)}
 \right)\\ &&+
\sum_{k=0}^{\infty}\frac{1}{k!} \left( \frac{\partial}{\partial \lambda} A^{\alpha
B_1\cdots B_{k}} \right) \tilde{l}_{B_1}\cdots \tilde{l}_{B_k} t^{\beta_1}\cdots t^{\beta_N} +\\
&& +\sum_{k=0}^{\infty}\frac{1}{k!} \left(\frac{\partial}{\partial \lambda_{ll}}A^{\alpha
B_1\cdots B_{k}}\right)\tilde{l}_{B_1}\cdots \tilde{l}_{B_k}
h^{(\beta_1\beta_2}t^{\beta_3}\cdots t^{\beta_{N)}}\begin{pmatrix}
  N \\
  2
\end{pmatrix} =\\
&& =\sum_{k=1}^{\infty}\frac{1}{(k-1)!} A^{\alpha B_1\cdots
B_{k-1}\beta_1\cdots\beta_N}\tilde{l}_{B_1}\cdots \tilde{l}_{B_{k-1}},
\end{eqnarray*}
where conditions $(\ref{20})$ have been used in the last passage. So we have proved that
derivation of eq. $(\ref{19})$ with respect to $l_{\beta_1\cdots \beta_N}$ is equivalent
to its derivation with respect to $\tilde{l}_{\beta_1\cdots \beta_N}$, but considering
independent the components of this tensor, except for the symmetry. Proceeding with the
subsequent derivatives and calculating the result at equilibrium, we find eq.
$(\ref{19a})$. In other words we can forget eq. $(\ref{19a})$ but we have to retain eqs.
$(\ref{20})$. We have then to transform eqs. $(\ref{10''})$, $(\ref{13})$ and
$(\ref{20})$ in conditions for the tensor $A^{\alpha B_1\cdots B_k}$; the above mentioned
symmetry of this tensor ensures that eq. $(\ref{10''})$ is satisfied. Before imposing
eqs. $(\ref{13})$ and $(\ref{20})$, we note that the most general expression for a
symmetric tensor depending on the scalars $\lambda$, $\lambda_{ll}$ and on $t^\alpha$ is
\begin{equation}\label{22}
A^{\alpha_1\alpha_2\cdots
\alpha_{Nk+1}}=\sum_{s=0}^{\left[\frac{Nk+1}{2}\right]}\begin{pmatrix}
  Nk+1 \\
  2s
\end{pmatrix} g_{k,2s}(\lambda,\lambda_{ll})h^{(\alpha_1\alpha_2}\cdots
h^{\alpha_{2s-1}\alpha_{2s}}t^{\alpha_{2s+1}}\cdots t^{\alpha_{Nk+1})}
\end{equation}
where the binomial factor has been introduced for later convenience. Thanks to this, eqs.
$(\ref{20})$ become
\begin{equation}\label{23}
  \begin{cases}
   g_{k+1,2s}=\frac{\partial}{\partial \lambda}g_{k,2s} \\
   \qquad \qquad \qquad \qquad \qquad \qquad \qquad \text{for }s=0,\cdots,\left[\frac{Nk+1}{2}\right]\\
   g_{k+1,2s+2}=\frac{2s+1}{2s+3}3\frac{\partial}{\partial
   \lambda_{ll}}g_{k,2s}.
  \end{cases}
\end{equation}
There remains to consider eq. $(\ref{13})$; thanks to eq. $(\ref{19})$, $(\ref{19z})$ and
$(\ref{22})$, its value at equilibrium is
\begin{equation*}
0=g_{0,0}+\frac{2}{3}\lambda_{ll}g_{1,2}
\end{equation*}
which, thanks to eq. $(\ref{23})_2$, becomes
\begin{equation*}
0=g_{0,0}+\frac{2}{3}\lambda_{ll}\frac{\partial}{\partial \lambda_{ll}}g_{0,0}.
\end{equation*}
Its solution is
\begin{equation}\label{28.a}
g_{0,0}=\lambda_{ll}^{-\frac{3}{2}}G_{0,0}(\lambda),
\end{equation}
with $G_{0,0}(\lambda)$ an arbitrary single variable function. \\
But eq. $(\ref{13})$ is equivalent to its value at equilibrium, and to its $r^{th}$
derivatives with respect to $l_{B_i}$ calculated at equilibrium, for all values of r. The
$r^{th}$ derivatives of eq. $(\ref{13})$ with respect to $l_{B_i}$ is
\begin{eqnarray}\label{*}
 0&=&\delta_j^\alpha \frac{\partial^r h^{'\mu} t_\mu}{\partial l_{B_1}\cdots\partial
l_{B_r}} + N \frac{\partial^{r+1}h{'^\alpha}}{\partial l_{B_1} \cdots \partial l_{B_r}
\partial l_{\alpha_1 \cdots \alpha_N}} t_{\alpha_N} l_{\alpha_1 \cdots \alpha_{N-1}j} +\nonumber\\
&+&N r t_{\alpha_N} \frac{\partial^{r}h^{'\alpha}}{\partial l_{\alpha_1 \cdots \alpha_N}
\partial l_{(B_1} \cdots \partial l_{B_{r-1}}} \frac{\partial l_{\alpha_1 \cdots
\alpha_{N-1}j}}{\partial l_{B_{r)}}},
\end{eqnarray}
where the indicated symmetrization is treated as the multi-index $B_i$ was a single
index. The eq. $(\ref{*})$ can be easily proved with the iterative procedure. Now we have
to calculate this expression at equilibrium. Let us evaluate each single term of this
relation.\\
$\bullet$ Thanks to eqs. $(\ref{19})$ and $(\ref{19z})$, we have for the first term
\begin{equation*}
\delta_j^\alpha \left(\frac{\partial^r h^{'\mu} t_\mu}{\partial l_{B_1}\cdots\partial
l_{B_r}}\right)_{eq}=\delta_j^\alpha A^{\mu B_1\cdots B_r}t_\mu.
\end{equation*}
$\bullet$ The second term at equilibrium, thanks to eq. $(\ref{19z})$, is
\begin{equation*}
\left(N \frac{\partial^{r+1}h^{'\alpha}}{\partial l_{B_1} \cdots \partial l_{B_r}\partial
l_{\alpha_1 \cdots \alpha_N}}t_{\alpha_N} l_{\alpha_1 \cdots \alpha_{N-1}j}\right)_{eq}=
N A^{\alpha B_1\cdots B_r \alpha_1\cdots \alpha_N}t_{\alpha_N} \frac{1}{3}\lambda_{ll}
\frac{2}{N}h_{j(\alpha_1}t_{\alpha_2}\cdots t_{\alpha_{N-1})}
\end{equation*}
The symmetrization in the right hand side can be omitted because the term is contracted
with a symmetric tensor. Now we use eq. $(\ref{22})$. We see that the terms containing
the factor $t^{\alpha_1}$ gives zero contribute, so that the above expression can be
written as
\begin{eqnarray*}
&&\sum_{s=1}^{\left[\frac{N(r+1)+1}{2}\right]}g_{r+1,2s}\begin{pmatrix}
N(r+1)+1\\
2s
\end{pmatrix}\frac{2s}{N(r+1)+1} h^{\alpha_1(\alpha_2}\cdots
h^{\alpha_{2s-1}\alpha_{2s}}t^{\alpha_{2s+1}}\cdots
t^{\alpha_{N(r+1)}}t^{\alpha)}t_{\alpha_N}\\
&& \cdot \frac{1}{3}\lambda_{ll}2h_{j \alpha_1}t_{\alpha_2}\cdots t_{\alpha_{N-1}}
\end{eqnarray*}
where the indexes in $B_1\cdots B_r$ and $\alpha_N$ are included into the $\alpha_i$;
after the contraction with $t_{\alpha_2} \cdots t_{\alpha_N}$ this expression becomes
\begin{equation*}
\sum_{s=1}^{\left[\frac{Nr+2}{2}\right]}\begin{pmatrix}
Nr+1\\
2s-1
\end{pmatrix}\frac{2}{3}\lambda_{ll} g_{r+1,2s}h_{j}^{(\gamma_2}\cdots h^{\gamma_{2s-1}\gamma_{2s}}t^{\gamma_{2s+1}}\cdots
t^{\gamma_{Nr+1}}t^{\alpha)}
\end{equation*}
where the indexes $\gamma^{\cdot}$ represent $B_1\cdots B_r$. \\
$\bullet$ Let us evaluate now the contribute of the last term in eq. $(\ref{*})$, i.e.
\begin{eqnarray*}
&&N r t_{\alpha_N} \left(\frac{\partial^{r}h^{'\alpha}}{\partial l_{\alpha_1 \cdots
\alpha_N}
\partial l_{B_1} \cdots \partial l_{B_{r-1}}}\right)_{eq}\frac{\partial l_{\alpha_1 \cdots
\alpha_{N-1} j}}{\partial l_{B_{r}}}=\\
 &&= N r t_{\alpha_N} A^{\alpha B_1\cdots
B_{r-1}\alpha_1 \cdots \alpha_N} g_{\alpha_1}^{(\beta_1^r} \cdots
g_{\alpha_{N-1}}^{\beta_{N-1}^r} h_j^{\beta_N^r)}
 = N r t_{\alpha_N} A^{\alpha \alpha_N B_1\cdots B_{r-1}(\beta_1^r\cdots \beta_{N-1}^r}
h_j^{\beta_N^r)}
\end{eqnarray*}
where we have esplicitated $B_r=\beta_1^r\cdots \beta_N^r$. We can now prove that
\begin{equation*}
N r t_{\alpha_N} \left(\frac{\partial^{r}h^{'\alpha}}{\partial l_{\alpha_1 \cdots
\alpha_N}
\partial l_{(B_1} \cdots \partial l_{B_{r-1}}}\right)_{eq}\frac{\partial l_{\alpha_1 \cdots
\alpha_{N-1} j}}{\partial l_{B_{r})}}
\end{equation*}
is symmetric with respect to two generic indexes $\beta_i^s$ and $\beta_q^t$, with $s\leq
t=1,\cdots ,r$. In fact it can be written as
\begin{eqnarray*}
&&\sum_{k=1}^{r} N t_{\alpha_N}\frac{\partial^r h^{'\alpha}}{\partial
l_{\alpha_1\cdots\alpha_N}\partial l_{B_1}\cdots\partial l_{B_{k-1}} \partial
l_{B_{k+1}}\cdots \partial l_{B_r}}\frac{\partial l_{\alpha_1\cdots\alpha_N j}}{\partial
B_k }= \\
&&=\sum_{k=1,\cdots,r}^{k\neq s, k\neq t} N t_{\alpha_N}A^{\alpha\alpha_N B_1\cdots
B_{k-1}
B_{k+1}\cdots B_r (\beta_1^k \cdots \beta_{N-1}^k}h_j^{\beta_N^k )}+\\
&&+N t_{\alpha_N}A^{\alpha\alpha_N B_1\cdots B_{s-1}
B_{s+1}\cdots B_{t-1} \beta_1^t \cdots \beta_{N}^t B_{t+1}\cdots B_r
(\beta_1^s \cdots \beta_{N-1}^s} h_j^{\beta_N^s )}+\\
&&+N t_{\alpha_N}A^{\alpha\alpha_N B_1\cdots B_{s-1} \beta_1^s \cdots \beta_{N}^s
B_{s+1}\cdots B_{t-1}B_{t+1}\cdots B_r (\beta_1^t \cdots \beta_{N-1}^t} h_j^{\beta_N^t
)}.
\end{eqnarray*}
The first of these terms is clearly symmetric with respect to $\beta_i^s$ and
$\beta_q^t$, while the sum of the last two is
\begin{eqnarray*}
&& t_{\alpha_N} A^{\alpha \alpha_N B_1 \cdots B_{s-1} B_{s+1}\cdots B_{t-1}\beta_1^t
\cdots \beta_{k}^t \cdots\beta_N^t B_{t+1}\cdots B_r \beta_1^s \cdots \beta_{i-1}^s
\beta_{i+1}^s \cdots \beta_N^s} h_j^{\beta_i^s}+\\
&&+t_{\alpha_N} A^{\alpha \alpha_N B_1 \cdots B_{s-1}\beta_1^s \cdots \beta_i^s \cdots
\beta_N^s B_{s+1}\cdots B_{t-1} B_{t+1}\cdots B_r \beta_1^t \cdots \beta_{k-1}^t
\beta_{k+1}^t \beta_N^t} h_j^{\beta_k^t} + \\
&& +\text{terms like }t_{\alpha_N} A^{\alpha \alpha_N \beta_i^s \cdots \beta_k^t \cdots}
h_j^{\cdot}
\end{eqnarray*}
that is obviously symmetric with respect to $\beta_i^s$ and $\beta_k^t$. \\
Consequently our tensor is symmetric with respect to every couple of indexes taken
between $B_1 \cdots B_r$, so that it can be expressed as
\begin{eqnarray}\label{p.17}
&&N r t_{\alpha_N}
A^{\alpha\alpha_N(\beta_1^1\cdots\beta_N^1\cdots\beta_1^r\cdots\beta_{N-1}^r}h^{\beta_N^r)}_j
= \sum_{s=0}^{\left[\frac{Nr}{2}\right]}2s\begin{pmatrix}
Nr\\
  2s
\end{pmatrix} g_{r,2s}h^{\alpha(\gamma_2}\cdots
h^{\gamma_{2s-1}\gamma_{2s}}t^{\gamma_{2s+1}}\cdots t^{\gamma_{Nr}}\cdot\nonumber\\
&&h^{\gamma_{Nr+1})}_j
+\sum_{s=0}^{\left[\frac{Nr}{2}\right]}\left(Nr-2s\right)\begin{pmatrix}
Nr\\
  2s
\end{pmatrix} g_{r,2s}t^\alpha h^{(\gamma_2\gamma_3}\cdots
h^{\gamma_{2s}\gamma_{2s+1}}t^{\gamma_{2s+2}}\cdots t^\gamma_{Nr}h^{\gamma_{Nr+1})}_j
\end{eqnarray}
Here we have calculated firstly $t_{\alpha_N}
A^{\alpha\alpha_N\beta_1^1\cdots\beta_N^1\cdots\beta_1^r\cdots\beta_{N-1}^r}$ by using
eq. $(\ref{22})$ and then distinguishing the terms in which $\alpha$ is index of an
$h^{..}$ from those in which it is an index of a $t^{.}$; finally we have multiplied the
result times $h_j^{\gamma_{Nr+1}}$ and symmetrized with respect to $\gamma_2\cdots\gamma_{Nr+1}$.\\
Until now we have finished to evaluate the three terms of eq. $(\ref{*})$ calculated at
equilibrium; so it becomes
\begin{eqnarray}\label{26}
&& 0=\sum_{s=1}^{\left[\frac{Nr+2}{2}\right]}g_{r+1,2s}\begin{pmatrix}
Nr+1\\
2s-1
\end{pmatrix}\frac{2}{3}\lambda_{ll}h_j^{(\gamma_2}\cdots h^{\gamma_{2s-1}\gamma_{2s}}t^{\gamma_{2s+1}}\cdots
t^{\gamma_{Nr+1}}t^{\alpha)}+\nonumber \\
&& + \sum_{s=1}^{\left[\frac{Nr+2}{2}\right]}g_{r+1,2s}\begin{pmatrix}
Nr+1\\
 2s-1
\end{pmatrix}\frac{2}{3}\lambda_{ll}h_j^{(\gamma_2}\cdots h^{\gamma_{2s-1}\gamma_{2s}}t^{\gamma_{2s+1}}\cdots
t^{\gamma_{Nr+1}}t^{\alpha)}+\nonumber \\
&&+ \sum_{s=0}^{\left[\frac{Nr}{2}\right]}2s\begin{pmatrix}
Nr\\
2s
\end{pmatrix}g_{r,2s}h^{\alpha(\gamma_2}\cdots h^{\gamma_{2s-1}\gamma_{2s}}t^{\gamma_{2s+1}}\cdots
t^{\gamma_{Nr}}h_j^{\gamma_{Nr+1})}
+ \nonumber \\
&& +\sum_{s=0}^{\left[\frac{Nr}{2}\right]}(Nr-2s)\begin{pmatrix}
Nr\\
2s
\end{pmatrix}g_{r,2s}t^{\alpha}h^{(\gamma_2\gamma_3}\cdots h^{\gamma_{2s}\gamma_{2s+1}}
t^{\gamma_{2s+2}}\cdots t^{\gamma_{Nr}}h_j^{\gamma_{Nr+1})}=\nonumber\\
&&=\sum_{s=0}^{\left[\frac{Nr}{2}\right]}(Nr+1)\begin{pmatrix}
Nr\\
2s
\end{pmatrix}g_{r,2s}h^{(\alpha\gamma_2}\cdots h^{\gamma_{2s-1}\gamma_{2s}}t^{\gamma_{2s+1}}\cdots
t^{\gamma_{Nr}}h_j^{\gamma_{Nr+1})}+\nonumber\\
&&+ \sum_{s=0}^{\left[\frac{Nr}{2}\right]}\begin{pmatrix}
Nr+1\\
2s+1
\end{pmatrix}\frac{2}{3}\lambda_{ll}g_{r+1,2s+2}h_j^{(\gamma_2}\cdots h^{\gamma_{2s+1}\gamma_{2s+2}}t^{\gamma_{2s+3}}\cdots
t^{\gamma_{Nr+1}}t^{\alpha)}
\end{eqnarray}
where in the second term we have changed the summation index s according to s=S+1.\\
Note that this equation is automatically symmetric. In $\cite{11}$ was proved that
$\frac{\partial \phi_{[k}}{\partial v_{i]}}=0$ is an identity for the case of 13 moments;
here we find that this property is valid also for an arbitrary number of
moments.\\
So we have proved that eq. $(\ref{*})$ amounts to
\begin{eqnarray}\label{32a}
0=(Nr+1)\begin{pmatrix}
  Nr \\
  2s
\end{pmatrix}g_{r,2s}+\begin{pmatrix}
  Nr+1 \\
2s+1
\end{pmatrix}\frac{2}{3}\lambda_{ll}g_{r+1,2s+2} \quad \text{, i.e.,}\nonumber\\
g_{r,2s}+\frac{2}{3}\lambda_{ll}\frac{1}{2s+1}g_{r+1,2s+2}=0 \quad \text{for }
s=0,\cdots,\left[\frac{Nr}{2}\right].
\end{eqnarray}
Consequently, all our conditions are equivalent to the scalar eqs. $(\ref{23})$,
$(\ref{28.a})$ and $(\ref{32a})$ which are constraints on the scalars $g_{r,2s}$ of the
expansion $(\ref{22})$. It remains to exploit them. For
$s=0,\cdots,\left[\frac{Nr}{2}\right]$ we can substitute $g_{r+1,2s+2}$ from eq.
$(\ref{23})$ into eq. $(\ref{32a})_2$ which now becomes
\begin{equation}\label{27}
\frac{2}{2s+3}\lambda_{ll}\frac{\partial }{\partial \lambda_{ll}}g_{r,2s}+g_{r,2s}=0
\end{equation}
whose solution is
\begin{equation}\label{28}
g_{r,2s}=\lambda_{ll}^{-\frac{2s+3}{2}}G_{r,2s}(\lambda)\quad \text{for
}s=0,\cdots,\left[\frac{Nr}{2}\right].
\end{equation}
In this way eq. $(\ref{23})_2$ is exausted, except for $s=\frac{Nr+1}{2}$ but only for
the case with Nr odd. \\
If $Nr$ is even eq. $(\ref{28})$ holds for all $g_{r,2s}$, while if $Nr$ is odd the
validity of eq. $(\ref{28})$ is not still proved for $g_{r,Nr+1}$. But for $Nr$ odd we
can use eqs. $(\ref{23})$ with $k=r$, $s=\frac{Nr+1}{2}$, i.e.,
\begin{equation}\label{A}
 \begin{cases}
 \frac{\partial}{\partial\lambda}g_{r,Nr+1}=g_{r+1,Nr+1}\\
 \frac{\partial}{\partial\lambda_{ll}}g_{r,Nr+1}=\frac{Nr+4}{Nr+2}\frac{1}{3}g_{r+1,Nr+3}.
 \end{cases}
\end{equation}
In the right hand sides we can use eq. $(\ref{28})$ because
$\frac{Nr+1}{2}\leq\left[\frac{N(r+1)}{2}\right]$ and
$\frac{Nr+3}{2}\leq\left[\frac{N(r+1)}{2}\right]$ hold, except for the trivial cases
N=1,2. In this way the system $(\ref{A})$ becomes
\begin{equation}\label{AA}
 \begin{cases}
 \frac{\partial}{\partial\lambda}g_{r,Nr+1}=\lambda_{ll}^{-\frac{Nr+4}{2}}G_{r+1,Nr+1}(\lambda)\\
 \frac{\partial}{\partial\lambda_{ll}}g_{r,Nr+1}=\frac{Nr+4}{Nr+2}\frac{1}{3}\lambda_{ll}^{-\frac{Nr+6}{2}}G_{r+1,Nr+3}(\lambda).
 \end{cases}
\end{equation}
The integrability conditions for this system gives
\begin{equation}\label{ic}
G'_{r+1,Nr+3}= \frac{-3}{2}(Nr+2)G_{r+1,Nr+1}.
\end{equation}
After that the system $(\ref{AA})$ can be integrated and gives
\begin{equation}\label{solution}
g_{r,Nr+1}=\lambda_{ll}^{-\frac{Nr+4}{2}}G_{r,Nr+1}(\lambda)+c_{r,Nr+1},
\end{equation}
with
\begin{equation}\label{definition}
G_{r,Nr+1}=-\frac{2}{3}\frac{1}{Nr+2}G_{r+1,Nr+3},
\end{equation}
while $c_{r,Nr+1}$ is an arbitrary constant arising from integration. So eq. $(\ref{28})$
is a valid so\-lu\-tion also in the case $Nr$ odd and $s=\frac{Nr+1}{2}$, except to add
the
arbitrary constant $c_{r,Nr+1}$.\\
Now we can see that this constant doesn't occur in eq. $(\ref{23})_1$ (because the right
hand side is differentiated, while in the left hand side and in the case N(k+1) odd, we
have $2s\leq 2\left[\frac{Nk+1}{2}\right]$ from which $2s < N(k+1)+1$). Nor it occurs in
eqs. $(\ref{28.a})$, $(\ref{28})$, $(\ref{ic})$, $(\ref{definition})$ and $(\ref{32a})$
(the proof for this last equation amounts to verify that $ \left[\frac{Nr}{2}\right]<
\left[\frac{Nr+1}{2}\right]$ for Nr odd and $\left[\frac{Nr}{2}\right]+1<
\left[\frac{N(r+1)+1}{2}\right]$ for N(r+1) odd; obviously, in both of them we have N
odd. If r is odd too, we have to verify only the first one, i.e. $\frac{Nr-1}{2}<
\frac{Nr+1}{2}$, which is an identity; if r is even, we have to
verify only the second one, i.e. $\frac{Nr}{2}+1< \frac{N(r+1)+1}{2}$ which is true, at least for $N>1$).\\
On the other hand, the contribute of this constant to the tensor
$A^{\alpha_1\alpha_2\cdots \alpha_{Nk+1}}$ is $h^{(\alpha_1\alpha_2}\cdots h^{\alpha_{N_k}\alpha_{N_k+1})}\cdot$\\
$c_{k,Nk+1}$, as it can be seen from eq. $(\ref{22})$. The contribute of all these
constants to $h^{'\alpha}$ follows from eq. $(\ref{19})$ and reads
\begin{equation}\label{42}
\sum_{r=0}^{\infty}\frac{1}{(2r+1)!}c_{2r+1,N(2r+1)+1}h^{\alpha(\beta_1^1}\cdots
h^{\beta^1_{N-1}\beta^1_N}\cdots h^{\beta_N^{N-1}\beta_1^N}\cdots
h^{\beta^N_{N-1}\beta_N^N)}\cdot l_{\beta_1^1\cdots \beta_N^1}\cdots
l_{\beta_1^N\cdots\beta_N^N},
\end{equation}
where we have put $k=2r+1$.\\
It is easy to verify that this additional term satisfies identically the symmetry
conditions for eq. $(\ref{10''})$ and $(\ref{13})$ (in fact $t_{\alpha_N}$ is contracted
with an $h^{\alpha_N}$, for this additional term). In other words, we can assume eq.
$(\ref{28})$ for all $g_{r,2s}$ (also for $s=\left[\frac{Nr+1}{2}\right]$), except that,
in the case with N odd, we have to add to $h^{'\alpha}$ the additional term
$(\ref{42})$.\\
Let's then substitute from eq. $(\ref{28})$ into eq. $(\ref{23})_1$ and $(\ref{32a})$; so
they become
\begin{equation}\label{43}
G_{k+1,2s}=G'_{k,2s}\quad \text{for }s=0,\cdots,\left[\frac{Nk+1}{2}\right],
\end{equation}
\begin{equation}\label{44}
G_{r+1,2s+2}=-3\frac{2s+1}{2}G_{r,2s}\quad \text{for
}s=0,\cdots\,\left[\frac{Nr}{2}\right].
\end{equation}
But this last equation holds also for $s=0,\cdots,\left[\frac{Nr+1}{2}\right]$; this is
obvious when Nr is even, while it is just eq. $(\ref{definition})$ when Nr is odd
(remember that
we have eq. $(\ref{definition})$ only for the case with Nr odd).\\
After that, we see that eq. $(\ref{28.a})$ is contained in $(\ref{28})$ for r=s=0, while
eq. $(\ref{ic})$, by using eq. $(\ref{definition})$, becomes $G'_{N,Nr+1}=G_{r+1,Nr+1}$
which is just eq. $(\ref{43})$ with k=r and $s=\left[\frac{Nr+1}{2}\right]$ (remember
that eq.
$(\ref{ic})$ holds only for Nr odd).\\
There remain eqs. $(\ref{43})$ and $(\ref{44})$. To this end, let us define $H_{r,s}$
from
\begin{equation}\label{45}
G_{r,2s}=\left(\frac{-3}{2}\right)^r\frac{(2s)!}{2^s s!}H_{r,s}.
\end{equation}
In this way eqs. $(\ref{43})$ and $(\ref{44})$ become
\begin{equation}\label{46}
H_{r+1,s+1}=H_{r,s},\qquad H'_{r,s}=\frac{-3}{2}H_{r+1,s} \quad \text{for
}s=0,\cdots,\left[\frac{Nr+1}{2}\right].
\end{equation}
Eq. $(\ref{46})_1$ suggests to define $H_{r,s}$ also for $s>\left[\frac{Nr+1}{2}\right]$.
In fact, let h be a number such that $s+h\leq\left[\frac{N(r+h)+1}{2}\right]$ (for
example, $h=\left[\frac{2s-Nr+1}{N-2}\right]$); we can define $H_{r,s}=H_{r+h,s+h}$. In
this way eq. $(\ref{46})_1$ holds for all r and s. Regarding eq. $(\ref{46})_2$ we have
\begin{equation*}
H'_{r,s}=H'_{r+h,s+h}=\frac{-3}{2}H_{r+h+1,s+h}=\frac{-3}{2}H_{r+1,s};
\end{equation*}
in other word, also $(\ref{46})_2$ holds for all r and s. \\
After that,
\begin{itemize}
  \item if $r\geq s$ we have
\begin{equation}\label{47}
H_{r,s}=H_{r-s,0}=\left(\frac{-2}{3}\right)^{r-s}\frac{d^{r-s}H_{0,0}}{d\lambda^{r-s}}
\end{equation}
  \item if $r<s$ we have
\begin{equation}\label{48}
H_{r,s}=H_{0,s-r}.
\end{equation}
\end{itemize}
In this way $H_{r,s}$ is known except for $H_{0,p}$.\\
On the other hand, it is easy to see that $(\ref{47})$  and $(\ref{48})$ satisfy eq.
$(\ref{46})_1$. Regarding $(\ref{46})_2$, we see that
\begin{itemize}
  \item if $r\geq s~\Rightarrow~r+1\geq s$, we have to use eq. $(\ref{47})$ for both
  sides of eq. $(\ref{46})_2$ and it becomes an identity,
  \item if $r=s-1$, we have to use eq. $(\ref{48})$ for the left hand side of eq.
  $(\ref{46})_2$ and eq. $(\ref{47})$ for the right hand side. The result is
  $H'_{0,1}=\frac{-3}{2}H_{0,0}$,
  \item if $r<s-1$, we have to use eq. $(\ref{48})$ for both sides of eq. $(\ref{46})_2$
  which becomes $H'_{0,s-r}=\frac{-3}{2}H_{0,s-r-1}$.
\end{itemize}
In conclusion, $H_{0,0}$ is arbitrary  and $H_{0,p}$ is defined by
\begin{equation}\label{49}
H'_{0,p}=\frac{-3}{2}H_{0,p-1},
\end{equation}
except for a constant arising from integration. after that, eq. $(\ref{47})$ and
$(\ref{48})$ give all the other functions $H_{r,s}$.
\section{The kinetic approach}
Let us now search a solution, for conditions $(\ref{10''})$ and $(\ref{13})$, of the form
\begin{equation}\label{17}
h^{'\alpha}=\int F\left(l_{\beta_1\cdots\beta_N} c^{'\beta_1}\cdots
c^{'\beta_N}\right)c^{'\alpha} d\underline{c'}
\end{equation}
where F is an arbitrary single variable function; it is related to the distribution
function, but this relation doesn't affect the following considerations, so that we
choose to omit it.\\
The symmetry for the left hand side of eq. $(\ref{10''})$ is certainly ensured;
remembering that $c^{'0}=1$, eq. $(\ref{13})$ becomes
\begin{equation}\label{18}
0=\int \frac{\partial}{\partial c'_j}\left[ F\left(l_{\beta_1\cdots \beta_N}
c^{'\beta_1}\cdots c^{'\beta_N}\right)c^{'\alpha}\right] d\underline{c'}
\end{equation}
which is certainly true. The expansion of eq. $(\ref{17})$ with respect to equilibrium is
\begin{equation*}
h^{'\alpha}=\sum_{k=0}^{\infty} \frac{1}{k!}\int F^{(k)}\left(\lambda
+\frac{1}{3}\lambda_{ll}c^{'2}\right) c{'^\alpha} c^{'B_1} \cdots c^{'B_k} d
\underline{c'} \tilde{l}_{B_1} \cdots \tilde{l}_{B_k}
\end{equation*}
where  eqs. $(\ref{22a})$, $(\ref{22b})$ and the multi-index notation have been used.
Then we have obtained eq. $(\ref{19})$ with
\begin{equation}\label{7.1}
A^{\alpha B_1\cdots B_k}=\frac{\partial^k B^{\alpha B_1\cdots B_k}}{\partial \lambda_k},
\qquad   B^{\alpha B_1\cdots B_k}=\int F
\left[\lambda+\frac{1}{3}\lambda_{ll}c{'^2}\right] c^{'\alpha}c^{'B_1}\cdots c^{'B_k}
d\underline{c'}.
\end{equation}
It is easy to verify that eqs. $(\ref{20})$ are satisfied with this expression. The
integral in eq. $(\ref{7.1})_2$ can be calculated with a well known procedure. To reach
faster the result, let us consider the tensor
\begin{equation*}
B^{\beta_1\cdots\beta_s\beta_{s+1}\cdots \beta_r}h_{\beta_1}^{\gamma_1}\cdots
h_{\beta_s}^{\gamma_s}t_{\beta_{s+1}}\cdots t_{\beta_r}= \int F\left[\lambda+\frac{1}{3}
\lambda_{ll}c{'^2} \right]c^{'\beta_1}\cdots c^{'\beta_s} h_{\beta_1}^{\gamma_1} \cdots
h_{\beta_s}^{\gamma_s} d\underline{c'}.
\end{equation*}
The above tensor depends only on scalar quantities and is symmetric, so it is equal to
\begin{equation*}
  \begin{cases}
    0 & \text{if s is odd}, \\
    g_s(\lambda,\lambda_{ll})h^{(\gamma_1\gamma_2}\cdots h^{\gamma_{s-1}\gamma_s)} & \text{if s is even}.
  \end{cases}
\end{equation*}
To know $g_s(\lambda,\lambda_{ll})$ it suffices to multiply both members by
$h_{\gamma_1\gamma_2}\cdots h_{\gamma_{s-1}\gamma_s}$ obtaining
\begin{equation*}
\int_0^{\infty} F\left[\lambda+\frac{1}{3} \lambda_{ll}c{'^2} \right]
c^{'s+2}\left(\int_0^\pi \sin\theta d\theta \right) \left(\int_0^{2\pi}d\phi\right) dc' =
g_s(\lambda,\lambda_{ll})(s+1)
\end{equation*}
where we have changed the integration variables according to the rule
\begin{eqnarray*}
  &&c^{'1}=c' \sin\theta\cos\phi, \quad c^{'2}=c'\sin\theta\sin\phi, \quad c^{'3}=c'\cos\theta\\
  &&c'\in [0,+\infty[, \qquad \quad \theta\in[0,\pi],\qquad \qquad \phi\in[0,2\pi[.
\end{eqnarray*}
We obtain
\begin{equation*}
g_s(\lambda,\lambda_{ll})=\frac{4\pi}{s+1}\int_0^{\infty} F\left[\lambda+\frac{1}{3}
\lambda_{ll}c{'^2} \right] c^{'s+2} dc' =
\left(\lambda_{ll}\right)^{-\frac{s+3}{2}}G_s(\lambda)
\end{equation*}
with
\begin{equation}\label{7.2}
G_s(\lambda)=\frac{4\pi}{s+1}\int_0^{\infty} F\left[\lambda+\frac{1}{3} \eta^2 \right]
\eta^{s+2}d\eta, \qquad \eta=\sqrt{\lambda_{ll}}c'.
\end{equation}
For the sequel it will be useful to note that
\begin{eqnarray}\label{7.3}
&&G'_s(\lambda)=\frac{4\pi}{s+1}\int_0^{\infty}\left\{\frac{d}{d\eta}F\left[\lambda+\frac{1}{3}
\eta^2\right]\right\}\eta^{s+1}d\eta \frac{3}{2}=\nonumber\\
&&=4\pi\frac{-3}{2}\int_0^{\infty}F\left[\lambda+\frac{1}{3}\eta^2\right]
\eta^{s}d\eta =\nonumber \\
&&=\frac{-3}{2}(s-1) G_{s-2}
\end{eqnarray}
provided that $F\eta^{s+1}$ is infinitesimal for $\eta$ going to infinity. After that, we
have
\begin{eqnarray*}
B^{\gamma_1\cdots
\gamma_r}&=&B^{\beta_1\beta_2\cdots\beta_r}\left(h_{\beta_1}^{\gamma_1}+t_{\beta_1}t^{\gamma_1}\right)
\left(h_{\beta_2}^{\gamma_2}+t_{\beta_2}t^{\gamma_2}\right)\cdots
\left(h_{\beta_r}^{\gamma_r}+t_{\beta_r}t^{\gamma_r}\right)=\\
&=& \sum_{s=0}^{r}\begin{pmatrix}
 r\\
  s
\end{pmatrix}B^{\beta_1\cdots\beta_r}h^{(\gamma_1}_{\beta_1}\cdots
h^{\gamma_{s}}_{\beta_s}t_{\beta_{s+1}}t^{\gamma_{s+1}}\cdots
t_{\beta_r}t^{\gamma_r)}=\\
&=&\sum_{q=0}^{\left[\frac{r}{2}\right]}\begin{pmatrix}
 r\\
  2q
\end{pmatrix}h^{(\gamma_1\gamma_2}\cdots
h^{\gamma_{2q-1}\gamma_{2q}}t^{\gamma_{2q+1}}\cdots
t^{\gamma_r)}\lambda_{ll}^{-\frac{2q+3}{2}}G_{2q}(\lambda).
\end{eqnarray*}
This allows to rewrite eq. $(\ref{7.1})$ as
\begin{equation*}
A^{\alpha B_1\cdots B_k}=\frac{\partial^k B^{\alpha B_1\cdots B_k}}{\partial
\lambda^k}=\sum_{q=0}^{\left[\frac{kN+1}{2}\right]} \begin{pmatrix}
kN+1 \\
  _2q
\end{pmatrix}h^{(\gamma_1\gamma_2}\cdots
h^{\gamma_{2q-1}\gamma_{2q}}t^{\gamma_{2q+1}}\cdots
t^{\gamma_{kN}}t^{\alpha)}\lambda_{ll}^{-\frac{2q+3}{2}}G^{(k)}_{2q}(\lambda).
\end{equation*}
This result confirms eq. $(\ref{22})$ also in the kinetic case, but with
$g_{k,2s}(\lambda,\lambda_{ll})=\lambda_{ll}^{-\frac{2s+3}{2}}G_{2s}^{(k)}(\lambda)$, and
it is easy to see that these functions $g_{k,2s}$ satisfy eq. $(\ref{23})$, as
consequence of eq. $(\ref{7.3})$. Also eqs. $(\ref{28})$ and $(\ref{solution})$ are
confirmed, with $G_{r,2s}(\lambda)=G_{2s}^{(r)}(\lambda)$, $c_{r,Nr+1}=0$. In this way we
see that the additional term $(\ref{42})$ is not present in the kinetic approach.
Moreover, the matrix $H_{r,s}$ defined in eq. $(\ref{45})$ becomes, in this approach
\begin{equation*}
H_{r,s}=\left[\frac{-2}{3}\right]^r\frac{2^s s!}{(2s)!}G_{2s}^{(r)}(\lambda),
\end{equation*}
and eq. $(\ref{49})$ becomes a consequence of eq. $(\ref{7.3})$. But the constants
arising from integration of eq. $(\ref{49})$ are not present in the kinetic approach,
because all the functions $G_s(\lambda)$ are defined by $(\ref{7.2})$ in terms of the
single variable function F.
\section{On subsystems.}
We aim to obtain now the model with N-1 instead of N through the method of subsystems. To
this end we firstly need the relation between the 4-dimensional Lagrange multipliers and
the 3-dimensional ones. The first of these are defined by eq. $(\ref{2'})$, from which we
obtain
\begin{eqnarray*}
dH^{\alpha}&=&l_{\alpha_1\cdots\alpha_N}^NdM_N^{\alpha\alpha_1\cdots\alpha_N}
=l^{\alpha_1\cdots\alpha_N}_NdM_N^{\alpha\beta_1\cdots\beta_N}g_{\alpha_1\beta_1}\cdots
g_{\alpha_N\beta_N}=\\
&=&l^{\alpha_1\cdots\alpha_N}_NdM_N^{\alpha\beta_1\cdots\beta_N}(h_{\alpha_1\beta_1}+t_{\alpha_1}t_{\beta_1})\cdots
(h_{\alpha_N\beta_N}+t_{\alpha_N}t_{\beta_N})=\\
&=&\sum_{r=0}^{N}\begin{pmatrix}
  N \\
  r
\end{pmatrix}
l^{\alpha_1\cdots\alpha_r\alpha_{r+1}\cdots\alpha_N}_NdM_N^{\alpha\beta_1\cdots\beta_r\beta_{r+1}\cdots\beta_N}
h_{\alpha_1\beta_1}\cdots
h_{\alpha_r\beta_r}t_{\alpha_{r+1}}\cdots t_{\alpha_N}t_{\beta_{r+1}}\cdots t_{\beta_N}=\\
&=&\sum_{r=0}^{N} \lambda_{j_1\cdots j_r}^N dF_N^{\alpha j_1\cdots j_r}
\end{eqnarray*}
\begin{eqnarray}\label{56}
&&\text{with}\quad \lambda_{j_1\cdots j_r}^N =\begin{pmatrix}
  N \\
  r
\end{pmatrix}l^{\alpha_1\cdots\alpha_r\alpha_{r+1}\cdots\alpha_N}_N h_{\alpha_1j_1}\cdots
h_{\alpha_rj_r}t_{\alpha_{r+1}}\cdots t_{\alpha_N}\nonumber\\
&&\text{and}\quad F_N^{\alpha j_1\cdots j_r}=
M_N^{\alpha\beta_1\cdots\beta_r\beta_{r+1}\cdots\beta_N}h_{\beta_1}^{j_1}\cdots
h_{\beta_r}^{j_r}t_{\beta_{r+1}}\cdots t_{\beta_N}.
\end{eqnarray}
Eq. $(\ref{56})_1$ gives the 3-dimensional Lagrange multipliers in terms of the
4-dimensional ones. We have introduced the index N to remember that we are considering
the model with N as maximum order of moments. In this way it will be distinguished from
that with N-1 instead of N.\\
The inverse of eq. $(\ref{56})_1$ is
\begin{eqnarray}\label{57}
l^{\alpha_1\cdots\alpha_N}_N&=&l^{\beta_1\cdots\beta_N}_N g^{\alpha_1}_{\beta_1}\cdots
g^{\alpha_N}_{\beta_N}=l^{\beta_1\cdots\beta_N}_N
(h^{\alpha_1}_{\beta_1}+t^{\alpha_1}t_{\beta_1})\cdots
(h^{\alpha_N}_{\beta_N}+t^{\alpha_N}t_{\beta_N})=\nonumber\\
&=&\sum_{s=0}^{N}\begin{pmatrix}
  N \\
  s
\end{pmatrix}
l^{\beta_1\cdots\beta_s\cdots\beta_N}_N h^{(\alpha_1}_{\beta_1}\cdots
h^{\alpha_s}_{\beta_s}t^{\alpha_{s+1}}\cdots t^{\alpha_N)}t_{\beta_{s+1}}\cdots t_{\beta_N}=\nonumber\\
&=&\sum_{s=0}^{N} \lambda^{(\alpha_1\cdots\alpha_s}_N t^{\alpha_{s+1}}\cdots
t^{\alpha_N)}.
\end{eqnarray}
The model with N-1 instead of N can be obtained as subsystem of the above one by taking
\begin{eqnarray}\label{57a}
\lambda_N^{\alpha_1\cdots\alpha_N}&=&0, \nonumber \\
\lambda_N^{\alpha_1\cdots\alpha_s}&=&\lambda_{N-1}^{\alpha_1\cdots\alpha_s}\quad
\text{for } s=0,\cdots , N-1.
\end{eqnarray}
We have now to express these relations in terms of the 4-dimensional Lagrange
multipliers; to this end we see that
\begin{equation}\label{58}
l^{\alpha_1\cdots\alpha_N}_N=\sum_{s=0}^{N-1} \lambda^{(\alpha_1\cdots\alpha_s}_{N-1}
t^{\alpha_{s+1}}\cdots t^{\alpha_N)},
\end{equation}
while eq. $(\ref{56})_1$, with N-1 instead of N, is
\begin{equation}\label{59}
\lambda_{j_1\cdots j_r}^{N-1}=\begin{pmatrix}
  N-1 \\
  r
\end{pmatrix}l^{\alpha_1\cdots\alpha_r\alpha_{r+1}\cdots\alpha_{N-1}}_{N-1}h_{\alpha_1
j_1}\cdots h_{\alpha_r j_r} t_{\alpha_{r+1}}\cdots t_{\alpha_{N-1}}.
\end{equation}
Then, by substituting eq. $(\ref{59})$ in eq. $(\ref{58})$ we find
\begin{equation*}
l^{\alpha_1\cdots\alpha_N}_N=\sum_{s=0}^{N-1} t^{(\alpha_{s+1}}\cdots t^{\alpha_N}
h^{\alpha_1}_{\gamma_1}\cdots h^{\alpha_s)}_{\gamma_s}
l^{\gamma_1\cdots\gamma_s\gamma_{s+1}\cdots\gamma_{N-1}}_{N-1} t_{\gamma_{s+1}}\cdots
t_{\gamma_{N-1}}\begin{pmatrix}
  N-1 \\
  s
\end{pmatrix}
\end{equation*}
from which
\begin{equation}\label{60}
l^{\alpha_1\cdots\alpha_N}_N=l^{(\alpha_1\cdots\alpha_{N-1}}_{N-1}t^{\alpha_{N})},
\end{equation}
because
\begin{eqnarray*}
l^{\alpha_1\cdots\alpha_{N-1}}_{N-1}t^{\alpha_{N}}&=&l^{\gamma_1\cdots\gamma_{N-1}}_{N-1}
t^{\alpha_{N}}(h^{\alpha_1}_{\gamma_1}+t^{\alpha_1}t_{\gamma_1})\cdots
(h^{\alpha_{N-1}}_{\gamma_{N-1}}+t^{\alpha_{N-1}}t_{\gamma_{N-1}})=\\
&=&\sum_{s=0}^{N-1}\begin{pmatrix}
  N-1 \\
  s
\end{pmatrix}
l^{\gamma_1\cdots\gamma_s\cdots\gamma_{N-1}}_{N-1} h^{(\alpha_1}_{\gamma_1}\cdots
h^{\alpha_s}_{\gamma_s}t_{\gamma_{s+1}}t^{\alpha_{s+1}}\cdots
t_{\gamma_{N-1}}t^{\alpha_{N-1})}t^{\alpha_N}.
\end{eqnarray*}
Now, from eq. $(\ref{19z})$, we have
\begin{equation*}
l^{\alpha_1\cdots\alpha_{N-1}}_{N-1~~eq.}=\lambda t^{\alpha_1}\cdots t^{\alpha_{N-1}}
+\frac{1}{3}\lambda_{ll} h^{(\alpha_1\alpha_2}t^{\alpha_3}\cdots t^{\alpha_{N-1})}.
\end{equation*}
This and eq. $(\ref{19z})$ yield
\begin{equation*}
l^{\alpha_1\cdots\alpha_{N}}_{N~~eq.}=l^{(\alpha_1\cdots\alpha_{N-1}}_{N-1~~eq.}t^{\alpha_N)},
\end{equation*}
that is, eq. $(\ref{60})$ holds also when we calculate it at equilibrium. The deviation
of eq. $(\ref{60})$ from its value at equilibrium is
\begin{equation}\label{61}
\tilde{l}^{\alpha_1\cdots\alpha_{N}}_{N}=\tilde{l}^{(\alpha_1\cdots\alpha_{N-1}}_{N-1}t^{\alpha_N)};
\end{equation}
in other words, eq. $(\ref{60})$ holds when we substitute the Lagrange multipliers with
their deviation with respect to equilibrium. We can now substitute eq. $(\ref{61})$ into
eq. $(\ref{19})$; in this way we find the counterpart of $(\ref{19})$ with N-1 instead of
N. To this end we have to contract an index of each $B_1, \cdots, B_k$ with a $t_{.}$; in
other words, we have to contract the expression $(\ref{22})$ with
$t_{\alpha_{(N-1)k+2}}\cdots t_{\alpha_{Nk+1}}$. It is easy to verify that in this way
eq. $(\ref{22})$ remains unchanged except that now N-1 replaces N; obviously this is true
also for $g_{k,2s}$ where s now goes from 0 to $\left[\frac{(N-1)k+1}{2}\right]$. This
property is transferred to $G_{k,2s}$ for eq. $(\ref{28})$ and to $H_{r,s}$ for eq.
$(\ref{45})$. But $H_{r,s}$ is defined by eqs. $(\ref{47})$ and $(\ref{48})$ in terms of
$H_{0,p}$ which are determined by eq. $(\ref{49})$. Therefore, the family of constants
arising by integrating
eq. $(\ref{49})$, is inherited also by the subsystem.\\
We have only to notice that from eq. $(\ref{48})$ it follows that $H_{0,p}$ is useful for
$H_{r,r+p}$ which, for eq. $(\ref{45})$ is useful for $G_{r,2(r+p)}$. It follows that
$H_{0,p}$ is present in the subsystem when $r+p\leq\left[\frac{(N-1)r+1}{2}\right]$, that
is $p\leq\left[\frac{(N-3)r+1}{2}\right]$, while for the initial system was useful when
$p\leq\left[\frac{(N-2)r+1}{2}\right]$. Now, for a fixed value of p, it is always
possible to find r such that both of the previous inequalities are satisfied. The only
difference is that in the subsystem, $H_{0,p}$ occurs only in terms of higher order with
respect to equilibrium, than in the initial system. This is true, provided that $N>3$,
that is if neighter the system, nor the subsystem are the 10 moments model.\\
But what happens to the other family of constants, that is for the supplementary term
$(\ref{42})$?\\
If N is even, the model has not this term and, consequently, it cannot be inherited by
the subsystem.\\
If N is odd, this term is present; but when we substitute eq. $(\ref{60})$ in
$(\ref{42})$ we obtain zero because each $t_{.}$ is contracted with a projector $h^{..}$.
We expected this result because, with N odd, we have N-1 even in which case the term
$(\ref{42})$ is not present. We may conclude that the other family of constants, or the
supplementary term $(\ref{42})$, disappears in the subsystem. Only the other family of
constants is inherited.\\
This can be seen also from the following viewpoint: the family of constants arising from
integration of eq. $(\ref{49})$, in the case N=3, will perpetuate also for the subsequent
values of N; equivalently, we can say that the closure in the model with a generic $N>3$
is exactly determined in terms of that with N=3, except for the supplementary term
$(\ref{42})$.
\section{Conclusions.}
In this paper we have applied the method of Extended thermodynamics to the case of an
arbitrary but fixed number of moments. This case has already been developed in the
kinetic approach. Here we have considered the macroscopic approach and the costitutive
functions have been determined up to whatever order with respect to thermodynamical
equilibrium. In this way we have been able to find the exact general solution of the
conditions present in extended thermodynamics for the case of many moments. The results
founded in the kinetic case, as we expected to found, are a particular case of the
present ones. Moreover, we have introduced an innovative 4-dimensional notation that
simpliflies very much the form of the equations. Finally, the present results in the case
N=3 are equivalent to those found in \cite{12}.\\


\begin{thebibliography}{99}
\bibitem[10]{10} S. Pennisi, T. Ruggeri, A new method to exploit the entropy principle and the
galielan invariance in the macroscopic approach of extended thermodynamics, to be
published,
\end{thebibliography}
\end{document}